\documentclass[sigconf]{acmart}

\AtBeginDocument{%
 }

\copyrightyear{2024}
\acmYear{2024}
\setcopyright{rightsretained}
\acmConference[SIGCSE 2024]{Proceedings of the 55th ACM Technical Symposium on Computer Science Education V. 1}{March 20--23, 2024}{Portland, OR, USA}
\acmBooktitle{Proceedings of the 55th ACM Technical Symposium on Computer Science Education V. 1 (SIGCSE 2024), March 20--23, 2024, Portland, OR, USA}
\acmDOI{10.1145/3626252.3630856}
\acmISBN{979-8-4007-0423-9/24/03}

\makeatletter
\gdef\@copyrightpermission{
  \begin{minipage}{0.3\columnwidth}
   \href{https://creativecommons.org/licenses/by/4.0/}{\includegraphics[width=0.90\textwidth]{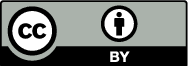}}
  \end{minipage}\hfill
  \begin{minipage}{0.7\columnwidth}
   \href{https://creativecommons.org/licenses/by/4.0/}{This work is licensed under a Creative Commons Attribution International 4.0 License.}
  \end{minipage}
  \vspace{5pt}
}
\makeatother

\usepackage{balance}
\usepackage[frozencache=true,cachedir=minted-cache]{minted}

\begin{document}

\title[Adding a Mistake-Based Familiarisation Step When Teaching Code Refactoring]{Fixing Your Own Smells: Adding a Mistake-Based Familiarisation Step When Teaching Code Refactoring}

 \author{Ivan Tan}
 \orcid{0009-0001-6300-5445}
 \affiliation{%
   \institution{Singapore Management University}
   \country{Singapore}
}
\email{ivantan@smu.edu.sg}

\author{Christopher M. Poskitt}
\orcid{0000-0002-9376-2471}
 \affiliation{%
   \institution{Singapore Management University}
   \country{Singapore}
}
\email{cposkitt@smu.edu.sg}

\begin{abstract}
    Programming problems can be solved in a multitude of functionally correct ways, but the quality of these solutions (e.g.~readability, maintainability) can vary immensely.
    When code quality is poor, symptoms emerge in the form of `code smells', which are specific negative characteristics (e.g.~duplicate code) that can be resolved by applying refactoring patterns.
    Many undergraduate computing curricula train students on this software engineering practice, often doing so via exercises on unfamiliar instructor-provided code.
    Our observation, however, is that this makes it harder for novices to internalise refactoring as part of their own development practices.
    In this paper, we propose a new approach to teaching refactoring, in which students must first complete a programming exercise constrained to ensure they will produce a code smell.
    This simple intervention is based on the idea that learning refactoring is easier if students are familiar with the code (having built it), that it brings refactoring closer to their regular development practice, and that it presents a powerful opportunity to learn from a `mistake'.
    We designed and conducted a study with 35 novice undergraduates in which they completed various refactoring exercises alternately taught using a traditional and our `mistake-based' approach, finding that students were significantly more effective and confident at completing exercises using the latter.
\end{abstract}

\ccsdesc[500]{Social and professional topics~Computing education}
\ccsdesc[300]{Software and its engineering~Maintaining software}

\keywords{Refactoring, code smells, code quality, software maintenance, software engineering, mistake-based learning, undergraduate course}

\maketitle

\section{Introduction}

Given any reasonably complex programming problem, there will be a multitude of functionally correct implementations that solve it.
Two solutions that always generate the intended outputs, however, are not always equally good.
Students are exposed to this fact early when they compare the performance of alternative solutions using recursion vs.~iteration, or quicksort vs.~bogosort~\cite{Gruber-et_al07a}.
Runtime performance, however, is not the only way that two solutions can differ: their \emph{code quality} can vary immensely too.
`Code quality' encompasses non-functional structural properties that arise from good engineering practices, e.g.~readability and maintainability~\cite{Sommerville15a,Borstler-et_al17a}.
In particular, a high-quality codebase is said to exhibit high cohesion and low coupling~\cite{Meyer97a}: functionally related elements are grouped together in modules, and those modules are sufficiently independent such that implementation changes in one should not cause another to inexplicably break.
When these principles are violated, concrete symptoms can emerge in the form of \emph{code smells}~\cite{Tufano-et_al15a}, which are characteristics (e.g.~the presence of duplicate code) that may indicate deeper problems.
In order to remove smells and improve code quality, software engineers apply \emph{refactoring patterns} that produce functionally equivalent but `odourless' code~\cite{Fowler18a}.

In undergraduate computing curricula, refactoring is typically introduced in software engineering modules taken after learning the fundamentals of programming.
A traditional delivery of the topic might teach a few examples of code smells, some corresponding refactoring patterns, and then challenge the students to apply them to some functionally correct (but smelly) instructor-provided code.
These exercises can be facilitated in a classroom or as part of an interactive online tutoring system~\cite{Keuning-et_al21a}.
While this mode of delivery has many advantages---the provided code is already working and simply needs refactoring---our own experiences have suggested that using instructor-provided code can make it harder for novices to internalise the concept into their own development practices.
This is because refactoring is introduced as a standalone exercise on someone else's code, rather than introducing it as a regular activity to be undertaken in any project they are developing.

In this paper, we propose a new approach to teaching code refactoring that embeds the concept as part of a multi-step exercise.
Students are first tasked to complete a programming exercise that is designed to ensure they will unwittingly produce smelly (but functionally correct) code.
The goal of this step is not to `bait' students, but to ensure that they are familiar with the code to be refactored.
Following this, they are taught to identify the smell that is present, and how to refactor it towards an odourless solution.
This simple intervention is based on three key ideas: (1)~that learning refactoring is simpler if students are already familiar with the targeted code, having written it themselves; (2)~that the approach aligns and embeds refactoring as part of their own coding practice; and (3)~that `mistakes'---in our case, induced code smells---present effective learning opportunities~\cite{Borasi94a}.

To assess the efficacy of this intervention, we conducted a study with 35 novice undergradate students.
We asked them to complete two groups of refactoring exercises for which the code smells were alternately taught using our `mistake-based' familiarisation approach or a traditional one.
We found that our approach led to significantly higher code smell identification and refactoring success rates, suggesting that students are able to apply the concepts more effectively.
The results encourage us to further explore mistake-based teaching approaches in other computing courses, e.g.~improving the learning opportunities in security courses by demonstrating the presence of security flaws in student code~\cite{Shar-Poskitt-et_al22a}.

\section{Related Work}

In an ITiCSE'17 working group report, B\"{o}rstler et al.~\cite{Borstler-et_al17a} analysed interviews with 34 students, educators, and developers on their perceptions of code quality.
They found that code quality was mostly understood in terms of indicators such as `readability', which are measures that code smells would score poorly against.
Notably, their interviewees ranked `education' lowest as the source they used most for learning about code quality, suggesting that undergraduate programmes can discuss the topic more thoroughly.
Effenberger and Pel\'{a}nek~\cite{Effenberger-Pelanek22a} buttress this point through their analysis of 114,000 functionally correct solutions in their introductory programming class, finding most of them to contain quality defects.

Bezerra et al.~\cite{Bezerra-Damasceno-Teixeira22a} conducted a study on the perceptions and challenges of undergraduate students when teaching code quality through code smell refactoring.
They highlighted a number of benefits, such as an improvement in problem solving and interpersonal skills, as well as a number of difficulties such as the fact that refactoring code can lead to new smells to further refactor.
They also observed that students found it more complicated to refactor smells when they struggled to interpret the source code---a problem our approach attempts to address by having students construct the code first in our familiarisation step.

Various techniques have been proposed for teaching code refactoring to undergraduates.
For instance, Haendler et al.~\cite{Haendler-et_al19a} developed an interactive web-based tutor in which smelly code is visualised in UML (`as-is') and students are challenged to refactor it towards a targeted design (`to-be').
The web-based tutor of Keuning et al.~\cite{Keuning-et_al20a,Keuning-et_al21a} allows students to request feedback which is generated based on some predefined rules provided by teachers, e.g.~rewrite steps.
In contrast, Haendler and Neumann~\cite{Haendler-Neumann19a} proposed designing `serious games' for teaching refactoring.
Students are presented with larger real-world code artefacts that are functionally correct but smelly, and are challenged to refactor them competitively.
Izu et al.~\cite{Izu-Denny-Roy22a} provide rules for simplifying conditional statements and some practice tasks to help students understand how to apply them.
In all these examples, smelly code is provided to students: our approach differs in that students build the smelly code, and thus can learn about refactoring using code that they are more familiar with.

Several existing tools can help to automatically identify smells~\cite{Fernandes-et_al16a} and assess the quality of code submitted in programming assignments.
Hyperstyle~\cite{Birillo-et_al22a}, for example, integrates code analysis tools into online educational platforms to provide feedback on readability, complexity, and patterns of repeated mistakes.
Prokic et al.~\cite{Prokic-et_al21a} integrate AI-based code quality assessment algorithms to identify issues as the code is written.
Our approach is similar in that we focus on the code the student is writing, but differs in that we induce a code smell intentionally to create a learning opportunity.

Many studies have demonstrated the effectiveness of learning from mistakes.
Borasi~\cite{Borasi94a} suggests that mistakes can be capitalised as a learning opportunity (or as `springboards' for inquiry).
Papert~\cite{Papert80a} views code debugging as such a learning opportunity, and our hypothesis is simply that these opportunities can be extended to mistakes in code quality too.
Ginat~\cite{Ginat08a} used erroneous solutions as a means to teach algorithm design: students would be introduced to an algorithm containing a common error, and would falsify inputs to trigger creative reasoning.
Ouh and Irawan~\cite{Ouh-Irawan18a} propose an experiential model for teaching software architecture, in which students undertake activities that simulate practical risks, helping them to learn how to identify, analyse, and resolve such risks in their own architectural solutions.
Shar et al.~\cite{Shar-Poskitt-et_al22a} demonstrated the value of security to web development students by introducing them to security scanners, and using them to uncover exploitable code in their own projects.
Griffin~\cite{Griffin19a} highlights the controversy of intentionally incorporating errors, but argues that cognitive psychology theories support the idea that intentional errors can promote learning.
Our work differs in that we focus on refactoring, and that a specific code quality `mistake' is induced in the familiarisation exercise (rather than provided directly by the instructor).

\section{Key Problems \& Research Questions}

Our context is Singapore Management University, where code smells and refactoring have been taught in a software engineering module taken by all undergraduate Information Systems students.
The delivery of this content has previously been `traditional', in that students were introduced to some key code smells~\cite{Fowler18a} and refactoring patterns~\cite{RefactoringGuru}, then were challenged to address the former using the latter in some simple instructor-crafted exercises.

We observed two key problems in this style of teaching.
First, especially for novice students, \emph{using instructor-provided code for refactoring exercises posed a familiarity barrier \textbf{(KP1)}}.
The issue we found was that for code beyond the very simplest, some students would struggle at the very first hurdle---familiarisation---and this would distract them from the primary learning objectives concerning smells and refactoring.
Second, we observed that students treated these as isolated exercises and were \emph{not gaining the confidence to apply refactoring to their own code \textbf{(KP2)}}.

Our proposed `mistake-based' intervention is inspired by the aforementioned works on learning from mistakes~\cite{Borasi94a,Papert80a,Ginat08a,Ouh-Irawan18a,Shar-Poskitt-et_al22a,Griffin19a}, as well as Abid et al.~\cite{Abid15a}, who observed benefits from asking students to `enhance' (add features to) existing code before asking them to refactor it.
In particular, rather than teach refactoring using instructor-provided code, we propose to first task students with completing a programming exercise that is designed to induce code that contains specific smells.
The idea is to ensure that students fully understand the code to be refactored (having written it themselves) and thus can separate out code familiarisation from code smell analysis in their learning.
Our intention is \emph{not} to `bait' students: we see this step as analogous to, for example, a programming exercise that first solves for specific inputs before guiding students to solve for all inputs, e.g.~by replacing conditionals with a loop.

The overall goal of this paper is to experimentally establish whether our `mistake-based' approach to teaching refactoring solves our key problems (KP1, KP2).
To guide our experiment design, we refined our goal to three research questions~(RQs):

\begin{itemize}
    \item[\textbf{RQ1:}] Which method results in a higher success rate for refactoring exercises?
    \item[\textbf{RQ2:}] Which method did students prefer and find more effective?
    \item[\textbf{RQ3:}] How confident are students at being able to identify code smells in the future?
\end{itemize}

RQ1 considers whether the introduction of a mistake-based familiarisation step helps students to complete refactoring exercises, thus addressing KP1.
RQ2 considers the students' subjective views, i.e.~which of the methods do they prefer and find more effective.
Finally, RQ3 addresses KP2, and considers whether students are confident that they can apply their refactoring skills in the future.

\section{Methodology}
In this section, we describe the experiment design, pre-/post-surveys, and participants of our study.

\subsection{Experiment Design}

\begin{figure}[!t]
\includegraphics[width=1\linewidth]{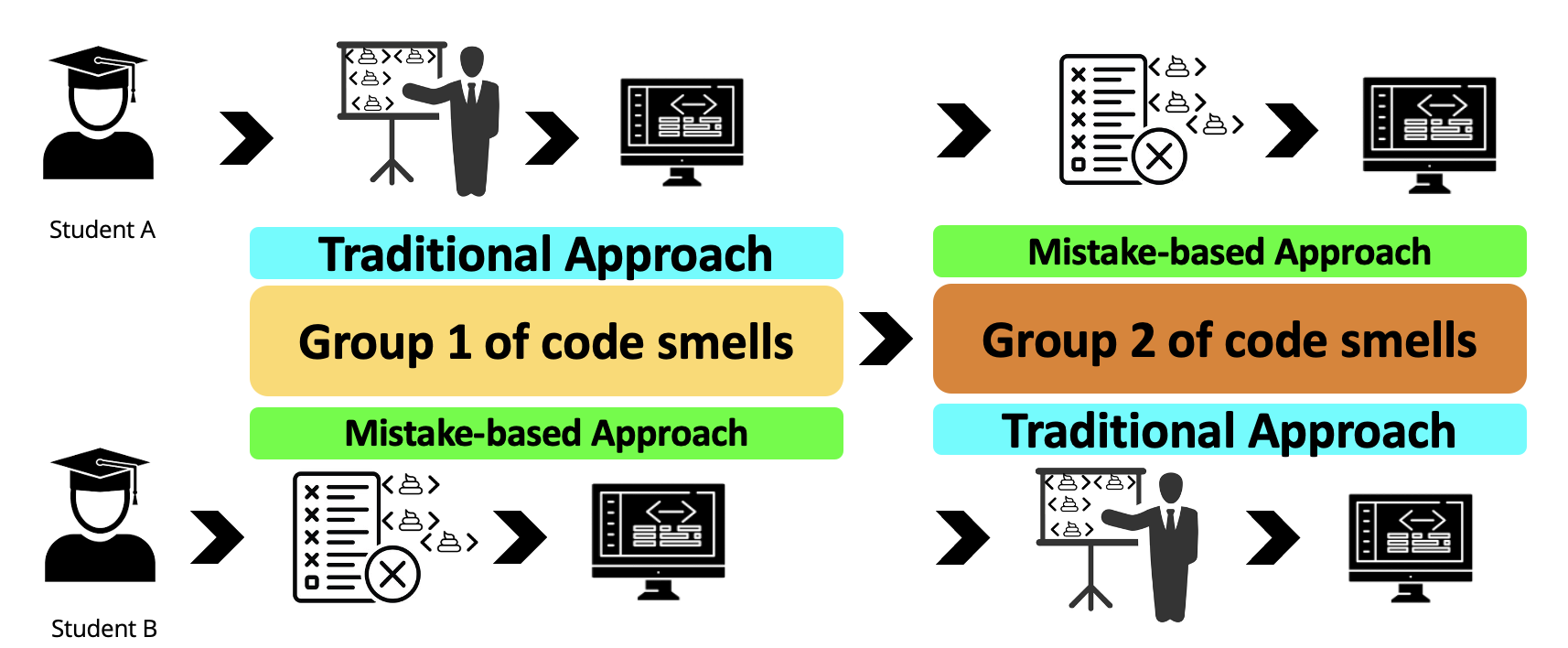}
\caption{High-level overview of the experiment protocol: participants are randomly allocated into two flows (A or B), and apply the two approaches to different code smell groups}
\label{fig:experiment_design}
\end{figure}

Figure~\ref{fig:experiment_design} presents a high-level overview of our experimental protocol, the detailed steps of which we describe in the following.
(The full set of exercises is also provided in our supplementary material~\cite{SupplementaryMaterials}.)
First, we defined two groups of code smells (three smells per group).
Group 1 contained Long Method, Long Parameter List, and Duplicate Code, whereas Group 2 contained Data Clumps, Large Class, and Primitive Obsession~\cite{Fowler18a}.
These were selected based on our expertise to ensure a roughly similar balance of difficulty and technical complexity between the two groups.

For each group, we prepared: (1)~a Python programming exercise with a template designed to induce some smelly code; and (2)~the smelly code the exercise is designed to induce.
Each exercise was set at a novice difficulty, given that the focus of the study was on code smells and not coding competency.
Group 1's exercise, for example, involved designing an object-oriented class for a sandwich shop (including, for example, a method for computing profit).
Listing~\ref{lst:coding_exercise_example} shows a snippet of the provided template (note that Lines~\ref{block:start}--\ref{block:end} are blank in the mistake-based approach; this is the `smelly' code we expect to induce). 
The template's pre-defined methods and parameter lists are designed to ensure solutions are largely similar (and in fact, the Long Parameter List code smell is guaranteed).

\begin{listing}
\begin{minted}[fontsize=\scriptsize,frame=single,obeytabs=true,tabsize=4,linenos,numbersep=-10pt,escapeinside=!!]{python}
    class Sandwich:
        def __init__(self):
            pass
    
        def calculate_profit(self, name, num_sold, recipe, price):
            profit = 0 # compute this below
            cost = 0 # add ingredient costs to this
            discount = 1 # change to 0.8 if num_sold >= 10
    
            # ENTER YOUR CODE BELOW
            for ingredient in recipe:!\label{block:start}!
                cost += ingredient_cost[ingredient]
            if num_sold >= 10:
                discount = 0.8
            price_per_sandwich = price * discount
            profit_per_sandwich = price_per_sandwich - cost
            profit = num_sold * profit_per_sandwich!\label{block:end}!
            # ENTER YOUR CODE ABOVE
            
            return round(profit, 2)
\end{minted}
  \caption{Coding exercise snippet. Lines~\ref{block:start}--\ref{block:end} show an example of the `smelly' code (long method) we expect the exercise to induce, and are blank in the mistake-based approach}
  \label{lst:coding_exercise_example}
\end{listing}

The study was conducted on a one-to-one basis with a research assistant: apart from conducting a briefing and pre-study survey (Section~\ref{sec:surveys}), the research assistant also provided oral instructions at all times.
Each student was randomly allocated into one of two flows: A or B.
Students in flow A were asked to apply the traditional approach to the Group 1 smells, followed by the mistake-based approach to the Group 2 smells.
Students in flow B, however, applied the mistake-based approach to Group 1 and the traditional approach to Group 2.
The specific steps of the approaches are described below.

For the traditional approach, we asked students to watch some short videos we produced~\cite{SupplementaryMaterials} that explain the code smells relevant to the Group using (different) instructor-provided code.
Each video introduces a specific code smell, explains why it occurs, why it should be refactored, how to refactor it, and then provides an example using instructor-provided code snippets.
This content is conveyed in under four minutes so that the videos remain bite-sized and engaging for the participants~\cite{Guo-Kim-Rubin14a}.
Afterwards, the research assistant provided the students with smelly code for the Group directly (e.g.~Listing~\ref{lst:coding_exercise_example} \emph{including} Lines~\ref{block:start}--\ref{block:end}).
In other words, the participants were asked to refactor instructor-provided smelly code based on what they learnt from the videos.

For the mistake-based approach, students were first asked to complete the (smell-inducing) programming exercise for Group 1 or 2 based on the template we provided (e.g.~Listing~\ref{lst:coding_exercise_example} \emph{excluding} Lines~\ref{block:start}--\ref{block:end}).
Again, the exercise and template were constrained to ensure that specific code smells from the given Group would emerge in the students' solutions.
In the event that a student struggled, the research assistant would provide hints to guide them towards the `smelly' solution, so as to ensure that the next part of the experiment would be able to carry on.
(We remark that hints were only provided in this familiarisation step, and not the refactoring exercise, which was the focus of the experiment.)
Following the coding exercise, the research assistant would provide a brief explanation of the code smells relevant to the Group (following the script used in the videos from the traditional approach), before asking the student to identify the smells in their own code.
For any smells the student failed to identify, the research assistant would record this before showing the student what they missed.
Finally, the research assistant gave another a brief oral explanation on the relevant refactoring patterns (again, following the video script), before asking the student to apply them where relevant to their own smelly code.

In both approaches, the research assistant recorded the number of smells identified and resolved by the participants.
They did not provide any help in identifying or resolving the smells until the participant indicated that they were finished.
To earn a point for code smell identification, they had to name the correct smell and locate where it was occurring.
For unidentified smells, these were highlighted by the research assistant to the students (with no point awarded), so that the refactoring part of the experiment could carry on.
To earn a point for resolving a smell, the refactored code had to remain correct (buggy misconceptions~\cite{Oliveira-Keuning-Jeuring23a} did not count) and the smell had to be removed to be considered successful.
Partial marks were not given for partial fixes.
At the end of the experiment, any unresolved smells were explained to the students by the research assistant for the participant's learning.

\subsection{Pre- and Post-Surveys}
\label{sec:surveys}
Prior to the experiment, we used a pre-study survey to collect some basic demographic information (e.g.~gender, year of study), as well as the pre-university institution they studied at (as some involve significant practical programming lessons).
We also asked them to rate their proficiency in Python, code smells, and code refactoring using Likert scales of 1--7 (based on the suggestion of~\cite{taherdoost2019best}).

\begin{figure}
\begin{minted}[fontsize=\tiny,frame=single,obeytabs=true,tabsize=4,numbersep=-10pt,escapeinside=!!]{text}
Q1.  Do you think your knowledge of code refactoring has improved? (Y/N)

Q2.  How familiar are you with the concept of code refactoring? (Likert)

Q3a. How confident are you in identifying code smells from Group 1? (Likert)

Q3b. How confident are you in identifying code smells from Group 2? (Likert)

Q4a. How confident are you in refactoring identified code smells from Group 1? (Likert)

Q4b. How confident are you in refactoring identified code smells from Group 2? (Likert)

Q5a. Do you understand the videos for the traditional method? (Y/N)

Q5b. Did you understand the content covered in the mistake-based approach? (Y/N)

Q6.  How effective was the traditional method? (Likert)

Q7.  How effective was the mistake-based approach? (Likert)

Q8.  Which method did you prefer? (Traditional/Mistake-based/No preference)

Q9.  What did you like/dislike about the traditional method? (Open)

Q10. What did you like/dislike about the mistake-based approach? (Open)
\end{minted}
\caption{Post-experiment survey (Likert scales are 7-point)}
\label{fig:post-exp_survey}
\end{figure}

After the experiment, participants were asked to complete a post-study survey (Figure~\ref{fig:post-exp_survey}) to determine whether they perceived an improvement in their confidence to identify and resolve code smells.
The final questions involved free text responses to collect some qualitative assessments from the participants.
All Likert scales consisted of 7 points, where 1 indicates least confidence/effectiveness, 4 is neutral, and 7 indicates most confidence/effectiveness.

\subsection{Participants}
We recruited 35 undergraduate Information Systems students from our institution.
Among these students, 30 were in the first or second year of their Bachelor's degree, whereas the others were in their third or final years; 17 reported their gender as female with the remaining 18 reporting as male.
Given that refactoring is only taught towards the end of their degrees, the vast majority of the participants can be considered novices in this topic.
This was further confirmed by the pre-study survey, in which the majority of them rated their familiarity with refactoring as either 1 or 2 (out of 7), while proficiency in Python varied from 1 to 5 (out of 7), with most falling in the lower range (1 to 3).

\section{Results \& Analysis}
In this section, we will analyse the results gathered from the experiments in accordance to the research questions defined.

In conducting our analysis, we primarily employed the Wilcoxon signed-rank test.
The test was used as the key point was the difference between the two methods for each paired measurement (one participant), so that we could obtain a p-value to interpret against our null and alternative hypotheses.
Whether it was comparing success rates, confidence, or effectiveness, they were all based on comparing the two different methods: traditional or mistake-based.

\subsection{RQ1: Refactoring Success Rates} 

For RQ1, we evaluated the success rates for the refactoring exercises, based on the methods that were used to teach the student.
For each participant, four values were collected: (1) number of code smells identified when taught with the traditional method; (2) number of code smells refactored when taught with the traditional method; (3) number of code smells identified when taught with the mistake-based method; and (4) number of code smells refactored when taught with the mistake-based method.
Wilcoxon signed-ranked tests were then conducted to compare the means of the variables: one for identification across the two methods, and one for refactoring across the two methods.
The null hypothesis is that there is no significant difference in the success rates between the two methods, with the alternative hypothesis being that there~is. 

\begin{figure}
\includegraphics[width=\linewidth]{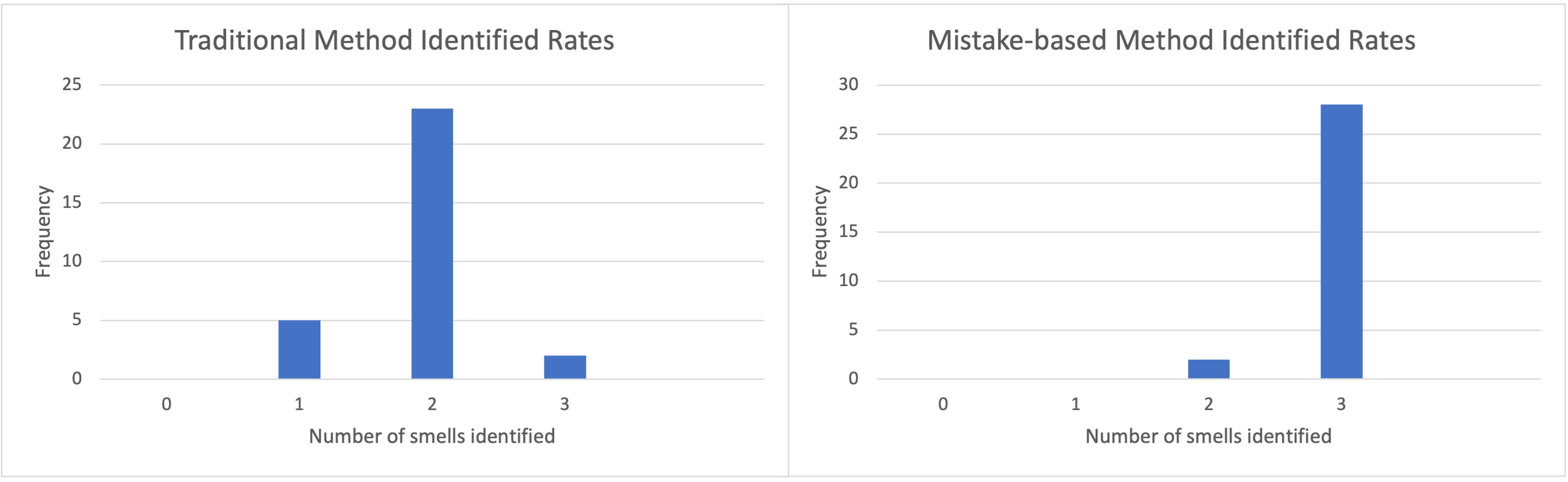}
\caption{Number of code smells identified (out of 3)}
\label{fig:identification_answer_rates}
\end{figure}

We begin with the code smell identification rate between the two methods.
Students were evaluated against a possible three code smells to be identified for each method, and only received the point if they were able to correctly identify what the code smell was and where it was located.
Figure~\ref{fig:identification_answer_rates} shows a comparison of the code smell identification success rates between the two methods.
We can see that for the mistake-based method, most students were able to identify all three code smells, whereas only a few were able to do so for the traditional method.
The average number of code smells identified for the mistake-based method was 2.93 (out of 3), whereas it was 1.9 for the traditional method.
The median for the mistake-based method was 3 (out of 3), whereas it was 2 for the traditional method.
The test returned a p-value of 3.2017e-06, and thus we accept the alternative hypothesis that there is a significant difference between the identification rates for the two methods.

\begin{figure}
\includegraphics[width=\linewidth]{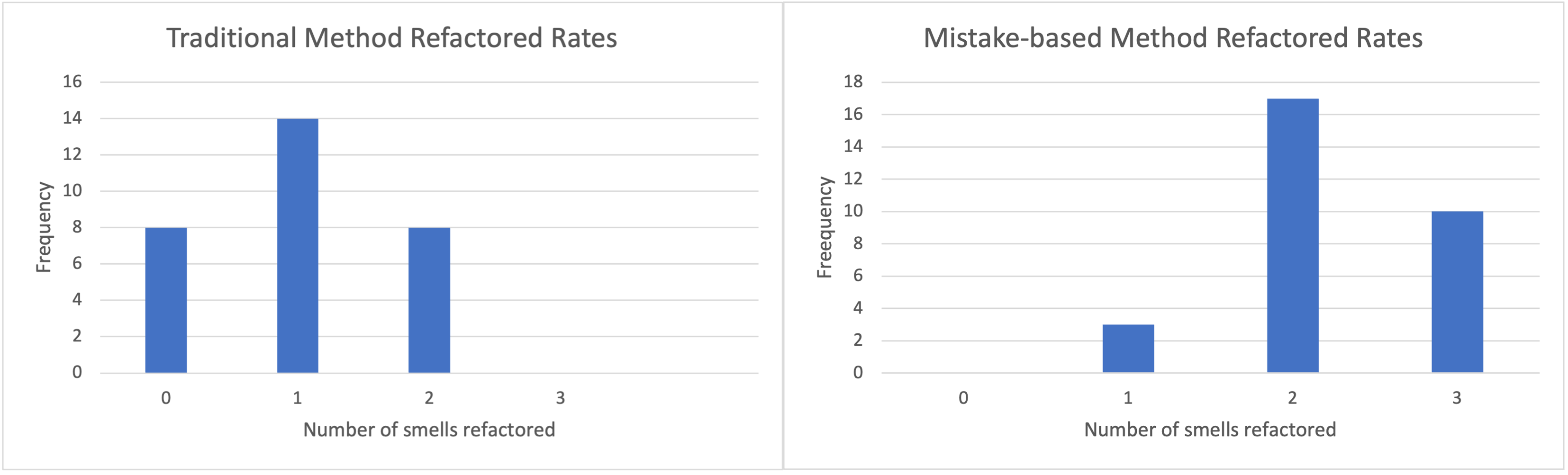}
\caption{Number of code smells successfully refactored (out of a possible 3)}
\label{fig:refactoring_answer_rates}
\end{figure}

Next, we look at the refactoring success rates.
Similar to identification, students were evaluated out of a possible three code smells to be refactored for each method.
They would only get the point if they were able to refactor the code smell and still maintain the function's logic.
For refactoring, the average number of code smells for the mistake-based method was 2.23 (out of 3), whereas it was 1 for the traditional method.
The median for the mistake-based method was 2 (out of 3), whereas it was 1 for the traditional method.
Figure~\ref{fig:refactoring_answer_rates} shows that for the mistake-based method, participants could refactor more code smells in general.
The test returned a p-value of 1.880441e-05, and thus we accept that there is a significant difference between the refactoring rates for the two methods.

To conclude the statistical analysis for RQ1, we can say that the mistake-based method was able to generate a higher success rate for both code smell identification and refactoring.
For code smells learnt with the mistake-based method, students were able to achieve a higher success rate than those learnt with the traditional method.
We believe that this is because the mistake-based approach removed the familiarity barrier for the exercises.

\subsection{RQ2: Preferred Method}
For RQ2, we used the post-study survey to establish which of the methods the students preferred learning with, and which they perceived to be more effective.
First, out of the 35 participants, 34 said that they preferred the mistake-based method, with the remaining participant having no preference.

Next, we want to find out which method the students perceived to be more effective.
Using Q6 and Q7 (Figure~\ref{fig:post-exp_survey}), we were able to gather quantitative data on a scale of 1--7 for the effectiveness ratings for both methods.
Again, our null hypothesis is that there is no significant difference between the effectiveness ratings for the two methods, whereas our alternative hypothesis is that there is.

\begin{figure}
\includegraphics[width=\linewidth]{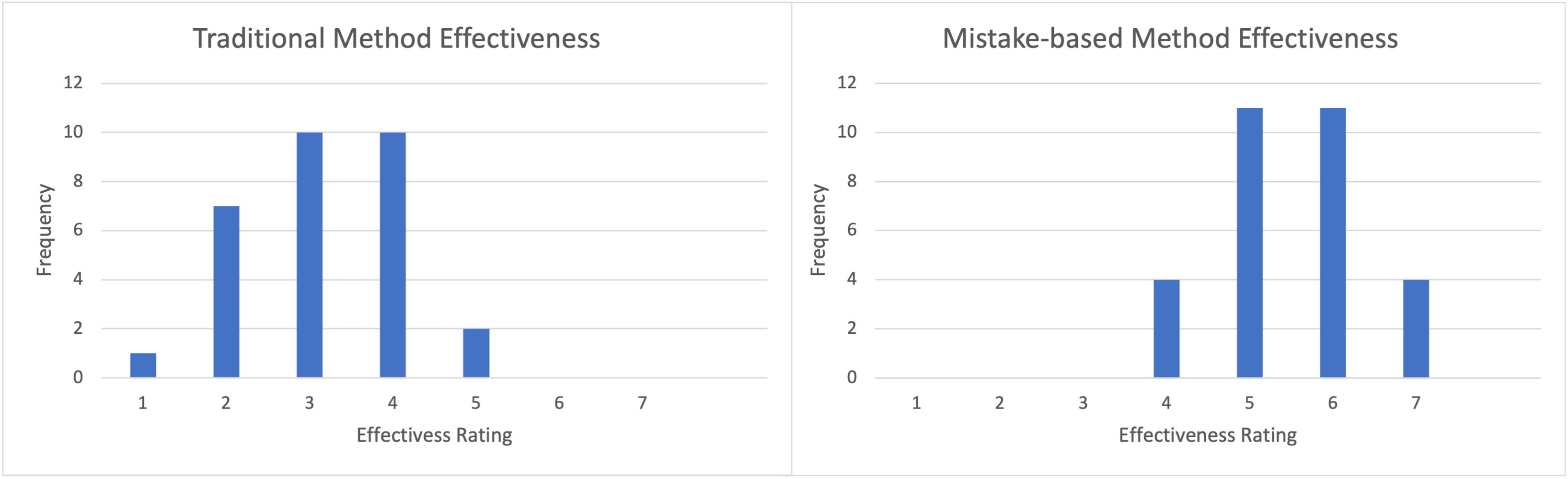}
\caption{How effective the students found the traditional vs.~mistake-based approaches (7-point Likert)}
\label{fig:effectiveness_ratings}
\end{figure}

From the histograms (Figure~\ref{fig:effectiveness_ratings}), we can see that the effectiveness rating for the traditional method is generally on the lower side, with its peak being a 4 (out of 7).
On the other hand, the mistake-based methods scored better in general.
The mistake-based method was able to get an average of 5.5 (out of 7), whereas the traditional approach obtained an average of 3.17.
The median for the mistake-based method was 5.5 (out of 7), whereas the traditional method obtained a median of 3.
With a p-value of 2.188529e-06, we thus accept the alternative hypothesis that there is a significant difference between the effectiveness ratings.

For RQ2, it appears that the preferred method in terms of effectiveness was the mistake-based method.
Based on our qualitative feedback from students (Q9 and Q10), we found out that they liked it more as it was a step-by-step approach, and they could digest the code as they built it.
Highlighting their mistakes was also crucial: students shared that it helped them understand the concept more.

\subsection{RQ3: Confidence at Refactoring}
For RQ3, we are interested in two things.
Firstly, whether the students were more confident in identifying code smells after the experiment.
Secondly, and most importantly, we want to establish whether they are more confident at identifying/resolving code smells learnt using the mistake-based or traditional method.

First, we look at the change in students' confidence before and after the experiment.
Data was obtained using pre- and post-study survey questions which required students to rate their confidence on a scale of 1--7.
Our null hypothesis was that there is no significant difference between the confidence levels before and after the experiment, with the alternative hypothesis being that there is one.

\begin{figure}
\includegraphics[width=\linewidth]{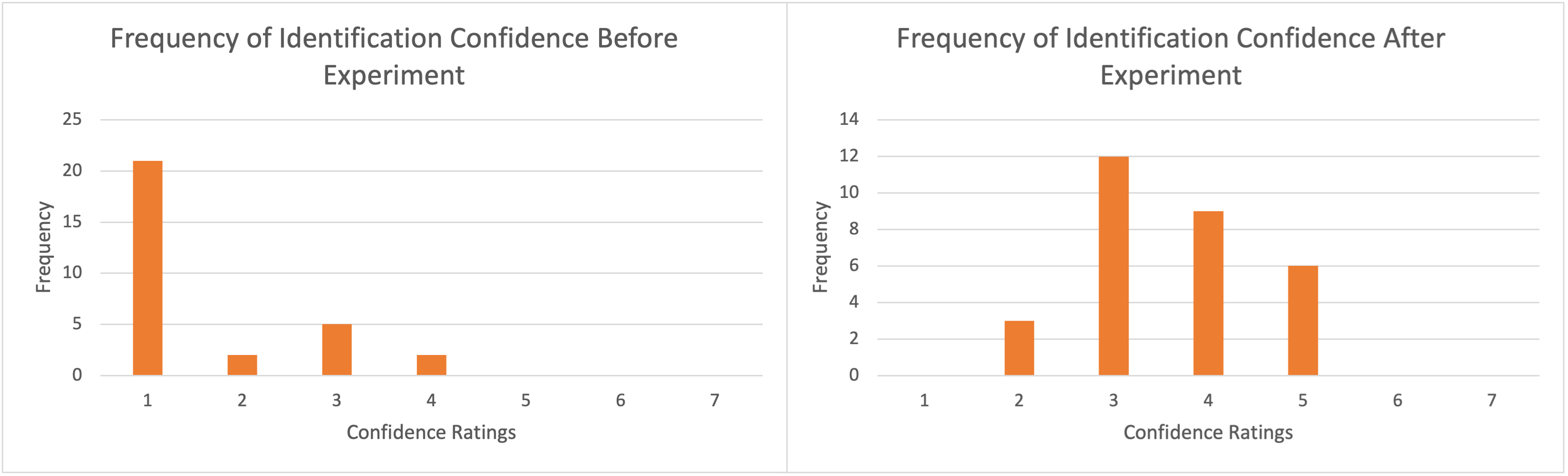}
\caption{Confidence identifying code smells before and after the study (7-point Likert)}
\label{fig:smell_ident_conf_befaft}
\end{figure}

We can see on the histograms (see Figure~\ref{fig:smell_ident_conf_befaft}) that there is a general increase in confidence after the experiment as compared to before. The average rating before was 1.6, which increased to an average of 3.6 after the experiment was conducted. Before the experiment, the median confidence was 1, and after the test, it increased to 3.5. The test returned a p-value of 1.597915e-06, allowing us to accept the alternative hypothesis, concluding that there is a difference between the confidence levels before and after the experiment.

Next, we want to evaluate whether students are more confident in identifying smells that they learnt using the mistake-based method versus those learnt using the traditional method.
Similar to RQ2, we will be using the identification confidence ratings that the students provided in our surveys.
Our null hypothesis is that there is no significant difference between the confidence levels between the two methods, with the alternative being that there is.

\begin{figure}
\includegraphics[width=\linewidth]{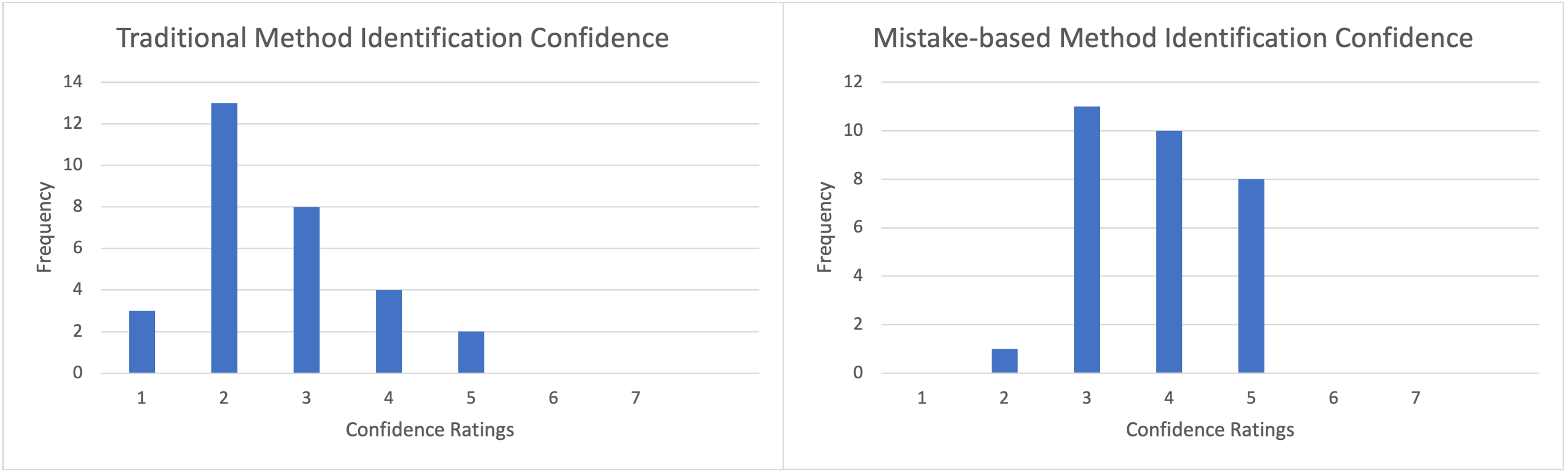}
\caption{Confidence identifying code smells learnt using the traditional vs.~mistake-based approach (7-point Likert)}
\label{fig:smell_ident_between_methods}
\end{figure}

From the histograms (see Figure~\ref{fig:smell_ident_between_methods}), we can observe that for the smells covered by the traditional method, the identification confidence level was lower than for the mistake-based method.
The average identification confidence level was 3.83 (out of 7) for the mistake-based method, but only 2.63 for the traditional method.
The median for the mistake-based method was 4 (out of 7), whereas the median for the traditional method was 2.
With a p-value of 2.161186e-06, and we thus accept our alternative hypothesis that there is a significant difference between the identification confidence levels for code smells learnt for each method.

We note an increase in the students' confidence in identifying code smells, particularly for those that they learnt using the mistake-based method.
We believe, again, that this is due to removing the familiarity barrier, allowing them to focus their learning entirely on refactoring instead of code comprehension.

\subsection{Threats to Validity}
Finally, we remark on some threats to the validity of our results.
First, the study was limited to undergraduates from a single institution.
It is possible that the results may not generalise due to differences in curricula, student profiles, and pedagogy.

Second, the exercises were designed around specific groupings of code smells and refactorings.
It is possible that the results will not apply to refactorings beyond those covered, or to groupings of smells/refactorings that do not maintain our difficulty balance.

Finally, to strive for objectivity in our evaluation, we used absolute values (0 or 1) to denote whether a student was successful in identifying or refactoring a certain code smell.
This might not be fully accurate, especially for refactoring, as there could be partially acceptable answers that we simply counted as 0.
For example, credit was not awarded for being able to correctly identify a code smell's location if the participant could not also name it.

\section{Reflections}
Some of the participants provided additional feedback at the end of the study.
A participant shared that the mistake-based approach was very similar to their experience of learning mathematics, where they would learn certain concepts better after getting the questions wrong first.
Another participant shared that they were inspired by the mistake-based approach, and would use it in their community service project involving teaching coding to secondary school children: they joked about asking the students to manually print ``Hello World!'' 20 times before introducing for-loops to them.
One participant also likened the experience to their internship at a startup, where one of their first code commits went through heavy code refactoring by their colleague, teaching them a ``lesson they would never forget''.
It was encouraging to hear these anecdotes, and it helped to validate our approach in ways we did not expect.

While this study has conveyed the potential value of a mistake-based methodology, challenges remain for practitioners to apply it in a classroom setting.
When setting the coding exercises in our study, we found it difficult to create questions where there was a balance between right and wrong, with just enough space and opportunity for a student to commit a smell that we were expecting.
If the question's design was too narrow, it would have been too obvious.
On the other hand, if the question's design was too broad, we would be getting mistakes that are irrelevant to the learning objectives.
Through a lot of iterations and trials, we were able to achieve a balance for this study, but in terms of using this approach in the classroom, this would be an important point for educators to consider.
Further research could potentially try to find ways to automate or use AI in generating these exercises.

\section{Conclusion \& Future Work}
In this paper, we proposed an approach to teaching refactoring that incorporates a `mistake-based' familiarisation step.
In other words, rather than refactor unfamiliar instructor-provided code, students complete a programming exercise that leads to smelly (but familiar) code for them to refactor instead.
This simple intervention is based on the idea that: (1) students will learn refactoring more effectively if they are already familiar with the targeted code, having built it; (2) it shows them refactoring isn't just about fixing other people's code, but can be incorporated into their own development practice; and (3) it aligns with the well-understood notion that `mistakes' provide a strong opportunity for learning.
We presented a study comparing our mistake-based teaching approach with a traditional one, finding that students were significantly more effective and confident at completing exercises.

This teaching methodology could potentially be used for other software engineering courses, and this is something that we are eager to test as well.
For instance, in a basic SQL / database management course, it might be useful for students to see the wrong results returned or erroneous merged tables created, allowing them to understand what was wrong with their query from the mistakes they made.
Similarly, in a web development course, the importance of securing web applications could be conveyed to students by subjecting their code to various security scanners and highlighting any vulnerabilities~\cite{Shar-Poskitt-et_al22a}.

\section*{Acknowledgements}

We are grateful to the anonymous referees for their helpful feedback on drafts of this paper.
We are also grateful to Sun~Jun and Ouh~Eng~Lieh for their helpful comments during the `UResearch' programme at SMU.
Thanks, finally, to the many students who kindly spent some time to participate in this study.

\bibliographystyle{ACM-Reference-Format}


\end{document}